# Accurate finite-difference micromagnetics of magnets including RKKY interaction - analytical solution and comparison to standard micromagnetic codes

D. Suess[1,2], S. Koraltan[1], F. Slanovc[1], F. Bruckner[1], C. Abert[1,2]

[1] Faculty of Physics, University of Vienna, 1090 Vienna, Austria

[2] Platform MMM Mathematics–Magnetism–Materials, University of Vienna, 1090

**Abstract:** Within this paper we show the importance of accurate implementations of the RKKY interactions for antiferromagnetically coupled ferromagnetic layers with thicknesses exceeding the exchange length. In order to evaluate the performance of different implementations of RKKY interaction, we develop a benchmark problem by deriving the analytical formula for the saturation field of two infinitely thick magnetic layers that are antiparallelly coupled. This benchmark problem shows that state-of-the-art implementations in commonly used finite-difference codes lead to errors of the saturation field that amount to more than 20% for mesh sizes of 2 nm which is well below the exchange length of the material. In order to improve the accuracy, we develop higher order cell based and nodal based finite-difference codes that significantly reduce the error compared to state-of-the-art implementations. For the second order cell based and first order nodal based finite element approach the error of the saturation field is reduced by about a factor of 10 (2% error) for the same mesh size of 2 nm.

*1. Introduction*

Magnetic thin films build the backbone of various applications ranging from magnetic recording [1] to magnetic sensors [2]. Magnetic TMR and GMR sensors rely on antiferromagnetic coupled films in the reference system of the magnetic sensing system. In addition to the application of synthetic antiferromagnets in sensors, magnetic hard disk devices exhibit antiferromagnetically coupled soft magnetic underlayers, to minimize the magnetic strayfield. The coupling strength of these magnetic layers can be well tuned via RKKY interactions as an indirect magnetic exchange coupling through the 3d, 4d, and 5d transition metals, such as Ru or Cr [3], [4]. Other magnetic systems where antiferromagnetic coupling has to be described accurately are ferrimagnetic structures that may exhibit antiparallel coupling between the net magnetization of the ferrimagnet and an adjacent ferromagnetic layer [5].

Due to the importance of antiferromagnetic coupling, micromagnetic codes usually offer the possibility to describe RKKY interaction. Commonly this is achieved by scaling the bulk exchange interaction to represent the RKKY interaction [6], [7] .

The finite-difference method usually assumes a homogeneous magnetization in each simulation cell with the finite-difference sampling points chosen in the cell centers [6]–[9]. In contrast, finite-element micromagnetic codes discretize the magnetization with affine basis function with the unknown coefficients of the magnetization at the node points of the finite-element mesh. In between these node points the magnetization is commonly linearly interpolated [10]. The RKKY interaction can be implemented by adding a surface energy to the total energy, taking into account the RKKY interaction energy.

In Chapter 2 the micromagnetic theory including RKKY interaction is reviewed. It is shown that the RKKY interaction alters the boundary condition for the magnetization. In Chapter 3 two analytical test cases are presented. The first test case consists of infinitely thin antiferromagnetically coupled ferromagnetic layers. The second test case is the opposing limit of infinitely thick antiferromagnetically coupled ferromagnetic layers. As we show in this paper, commonly used finite-difference micromagnetic codes are able to accurately describe thin antiferromagnetic coupled layers. However, these codes introduce significant errors both in the equilibrium magnetization as well as in critical fieldsif perpendicularly inhomogeneous states are formed in thick magnetic layers..
In order to reduce these errors, we derive higher order implementations of the RKKY interaction of cell centered finite difference codes.

In Chapter 4. The higher order cell-based implementations and the node-based implementation are reviewed and compared to the result of the analytical benchmark problems presented in Chapter 3.

## 2. *General micromagnetic description of magnetic regions coupled via RKKY interaction*

In the following, we derive Figure 1 the differential equation for the magnetization in equilibrium from the total energy of the magnet shown in Figure 1. Here we assume the magnetic region $\Omega_1$ and $\Omega_2$ as well as the corresponding boundary conditions. Within the domains $\Omega_1$ and $\Omega_2$ the magnetization $\boldsymbol{m}_1(\mathbf{x})$ and $\boldsymbol{m}_2(\mathbf{x})$ are continuous functions in space. At the common interface $\partial\Omega_1 \cap \partial\Omega_2$ in general, the magnetization $\boldsymbol{m}_1(\mathbf{x})$ is not equal to $\boldsymbol{m}_2(\mathbf{x})$. Here, we assume a RKKY interaction between these magnetizations at the surfaces.

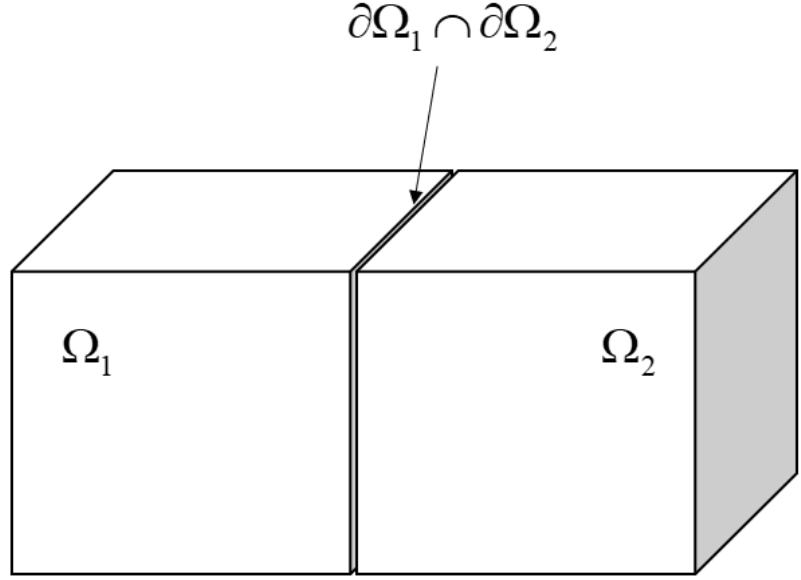

Figure 1: Considered domain of a magnet with region $\Omega_1$ and $\Omega_2$ that are coupled via the common interface $\partial\Omega_1 \cap \partial\Omega_2$.

For the considered general domain we start from the total energy $E_{tot}$ including exchange energy $E_{ex}$, anisotropy energy $E_{ani}$, Zeeman energy $E_{ex}$ and RKKY surface interaction $E_{rkky}$ in the form [11]:

$$E_{ani}(\mathbf{m}_1, \mathbf{m}_2) = -\int_{\Omega_1} K_{1,1} [\mathbf{m}_1 \mathbf{k}_1]^2 dV - \int_{\Omega_2} K_{1,2} [\mathbf{m}_2 \mathbf{k}_2]^2 dV \tag{2.1}$$

$$E_{ext}(\mathbf{m}_1, \mathbf{m}_2) = -\int_{\Omega_1} J_{s,1} \mathbf{m}_1 \mathbf{H}_{ext} dV - \int_{\Omega_2} J_{s,2} \mathbf{m}_2 \mathbf{H}_{ext} dV \tag{2.2}$$

$$E_{ex}(\mathbf{m}_1, \mathbf{m}_2) = \int_{\Omega_1} A_{1,x} [\nabla m_{x,1}]^2 + A_{1,y} [\nabla m_{y,1}]^2 + A_{1,z} [\nabla m_{z,1}]^2 dV +$$
$$\int_{\Omega_2} A_{2,x} [\nabla m_{x,2}]^2 + A_{2,y} [\nabla m_{y,2}]^2 + A_{2,z} [\nabla m_{z,2}]^2 dV \tag{2.3}$$

$$E_{rkky}(\mathbf{m}_1, \mathbf{m}_2) = -\int_{\partial\Omega_1 \cap \partial\Omega_2} J_{rkky} \mathbf{m}_1 \mathbf{m}_2 dA \tag{2.4}$$

The total energy is the sum of all energy terms

$$E_{tot}(\mathbf{m}_1, \mathbf{m}_2) = E_{ex}(\mathbf{m}_1, \mathbf{m}_2) + E_{rkky}(\mathbf{m}_1, \mathbf{m}_2) + E_{ext}(\mathbf{m}_1, \mathbf{m}_2) + E_{ani}(\mathbf{m}_1, \mathbf{m}_2) \tag{2.5}$$

The functional variation of the total energy according to Eq. (2.5) with respect to the magnetization $m_1$ and $m_2$ and the constrain $|\mathbf{m}_i| = 1$ is derived in detail in Appendix A. It leads to Browns equation within the volumes $\Omega_i$ for the equilibrium magnetization $m_i$:

$$\mathbf{m}_i \times \left[ 2\nabla \cdot (A_i \nabla \mathbf{m}_i) + J_{s,i} \mathbf{H}_{ext} + 2K_{1,i} (\mathbf{m}_i \mathbf{k}_i) \mathbf{k}_i \right] = 0 \tag{2.6}$$

With the well known exchange boundary condition

*Put A_i grad m_i n = 0 here*

(2.7)For the external interface $\partial(\Omega 1 \cup \Omega 2)$. $\partial\Omega_1 \cap \partial\Omega_2$ the following Robin boundary condition for the magnetization $m_1$ is obtained:

$$2\mathbf{A}_1 (\nabla \mathbf{m}_1) \mathbf{n} = -(\mathbf{m}_1 J_{rkky} \mathbf{m}_2) \mathbf{m}_1 + J_{rkky} \mathbf{m}_2 \tag{2.8}$$

An equivalent boundary condition is obtained for the magnetization $m_2$ for the common boundary $\partial\Omega_1 \cap \partial\Omega_2$.

For the case of $J_{rkky} = 0$ the boundary condition (2.8) reduces to the well-known boundary condition for the surface of the magnet,

$$\mathbf{A}_1 (\nabla \mathbf{m}_1) \mathbf{n} = 0 \tag{2.9}$$

For the case of $J_{rkky} = \infty$ it follows that $\mathbf{m}_1 = \mathbf{m}_2$ on $\partial\Omega_1 \cap \partial\Omega_2$. Hence, we obtain from Eq. (2.8) and the equivalent equation for the region $\Omega_2$ an equivalent right-hand side. It follows the well-known boundary condition between two magnets with different exchange constants,

$$2\mathbf{A}_1 (\nabla \mathbf{m}_1) \mathbf{n} - 2\mathbf{A}_2 (\nabla \mathbf{m}_2) \mathbf{n} = 0 \tag{2.10}$$

### 3. Analytic solutions for RKKY coupled layers

#### a. Limit for thin layers – homogenous magnetization within the layers

Here we consider the two antiparallelly coupled layers (i.e. $J_{rkky}<0$) that are sufficiently thin so that the magnetization remains perpendicularly homogenous within each layer. The thickness of each layer is *t*. The total energy for a system without anisotropy energy and demagnetizing energy is

$$E_{tot} = -J_{rkky}\int_A \mathbf{m}_1\mathbf{m}_2 dA - J_s\int_{\Omega_1}\mathbf{Hm}_1 dV - J_s\int_{\Omega_2}\mathbf{Hm}_2 dV \quad (2.11)$$

Assuming a homogeneous magnetization within each layerwe obtain

$$E_{tot}/F = -J_{rkky}\cos(2\varphi) - 2tJ_s H_x \cos(\varphi) \quad (2.12)$$

For a field in x-direction, where $\varphi$ is the angle between *x*-axis and magnetization.

The equilibrium angle is obtained from the solution of,

$$\frac{\partial E_{tot}}{\partial \varphi} = J_{rkky}\sin(2\varphi) + tJ_s H_x \sin(\varphi) = J_{rkky} 2\sin(\varphi)\cos(\varphi) + tJ_s H_x \sin(\varphi) = 0 \quad (2.13)$$

which leads to

$$\cos(\varphi) = m_{1,x} = m_{1,x} = -\frac{tJ_s H_x}{2J_{rkky}} \quad (2.14)$$

Obviously, the magnetization gets saturated for fields

$$H_{x,sat} = -\frac{2J_{rkky}}{tJ_s} \quad (2.15)$$

b. Limit for infinite thick layers – inhomogeneous magnetization within the layers

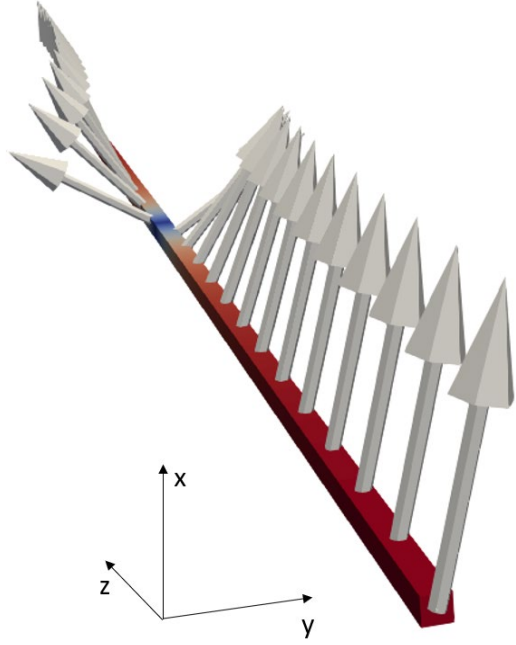

Figure 2: Considered domain of the 1d example. The region $\Omega_1$ is from $-\infty < z \leq 0$ and $\Omega_2$ is from $0 \leq z < \infty$. At the center at z=0 the spins are antiferromagnetically coupled. Without external field the magnetization points parallel to the y-axis which is parallel to the anisotropy field. Here a field is applied in the +x direction.

In order to test also RKKY implementations for the more complicated case, where perpendicularly inhomogeneous states are formed in the magnetic layer we design a benchmark problem with an analytical solution. We consider an infinite 1d magnet as shown in Figure 2. At z = 0 the spins are coupled antiferromagnetically due to RKKY interaction. We assume the anisotropy direction **k** = (0,1,0) and **H**$_{ext}$ = (H$_x$,0,0). As initial conditions we set for z > 0 the magnetization to $m_y$ = 1 and for z < 0 the magnetization to $m_y$ = -1. For vanishing external field this initial configuration is a stable state. For increasing H$_x$ the magnetization tends to align in x-direction while the anisotropy and RKKY interaction act against this alignment leading to inhomogeneous magnetization configurations within the ferromagnetic layers, see Fig. 2. Due to symmetry the magnetization can be written as,

$$\mathbf{m} = \begin{pmatrix} \cos(\varphi(z)) \\ \sin(\varphi(z)) \\ 0 \end{pmatrix} \quad (3.1)$$

where $\varphi(z)$ is the angle between the x-axis and the magnetization. At z=0 the angle $\varphi(z)$ shows in general a jump and it holds

$$\varphi(-z) = -\varphi(z) \tag{3.2}$$

As derived in Appendix B one obtains the relation between the angle $\varphi_0 = \varphi(z=0)$ at the RKKY interface and the applied field $H_x$,

$$H_x = \frac{2K_1}{J_s} h_x = \frac{2K_1}{J_s} \frac{1}{4} \frac{\cos(2\varphi_0) - 1 - \overbrace{\frac{J_{rkky}^2}{2AK_1}}^{j_{red}} \sin^2(2\varphi_0)}{\cos(\varphi_0) - 1} \tag{3.3}$$

$$j_{red} = \frac{J_{rkky}^2}{2AK_1} \tag{3.4}$$

The angle at the interface $\varphi_0$ as function of the external field $H_x$ for different values of $j_{red}$ obtained from the solution of Eq. (3.3) is shown in in *Figure 3* for different RKKY coupling strengths $j_{red}$.

After calculating the angle $\varphi_0$ for a particular field $H_x$, also the entire domain wall profile, $\varphi(z)$ can be obtained using Eq. (3.82). The analytic solution agrees very well with the finite-element solution of magnum.fe as shown in Figure 4.

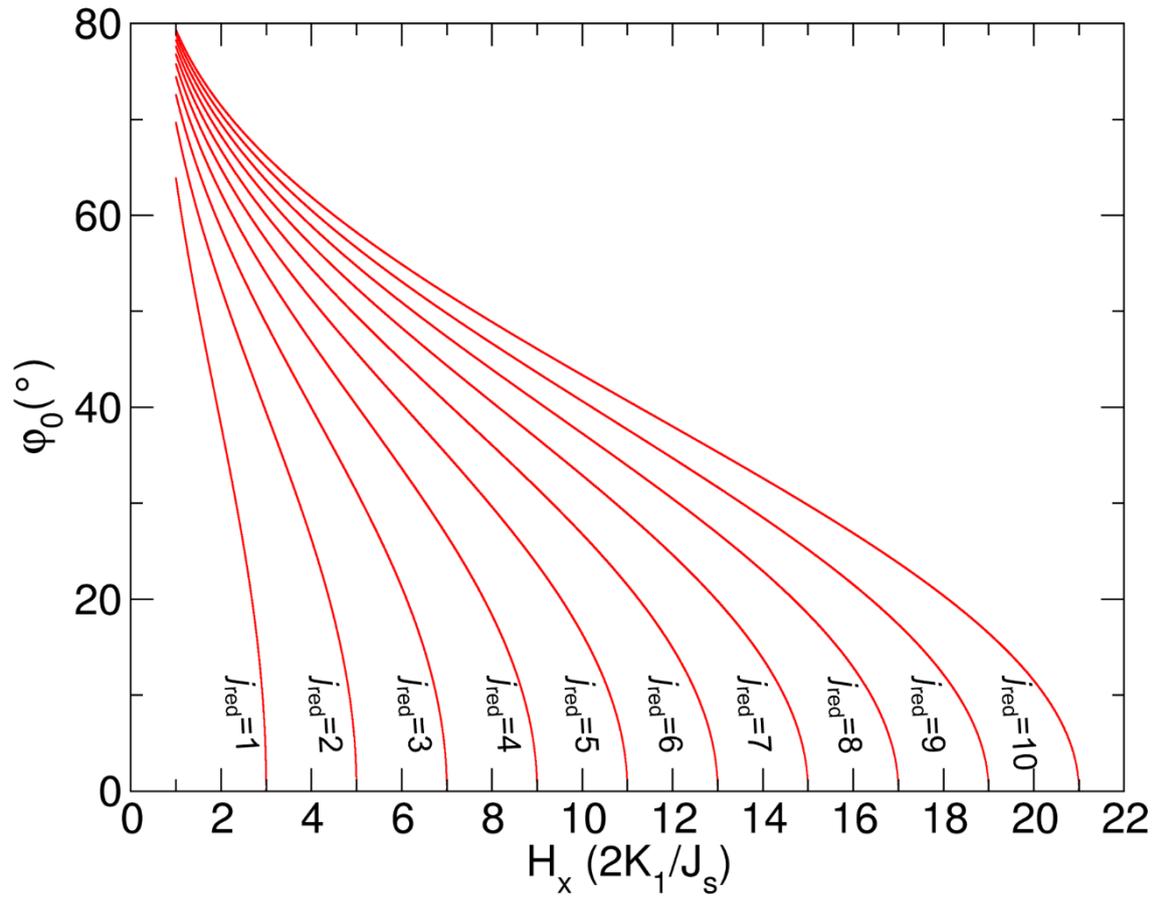

*Figure 3: Angle at the interface $\varphi_0$ as function of the external field $H_x$ for different values of $j_{red}$ obtained from the solution of Eq.* (3.3).

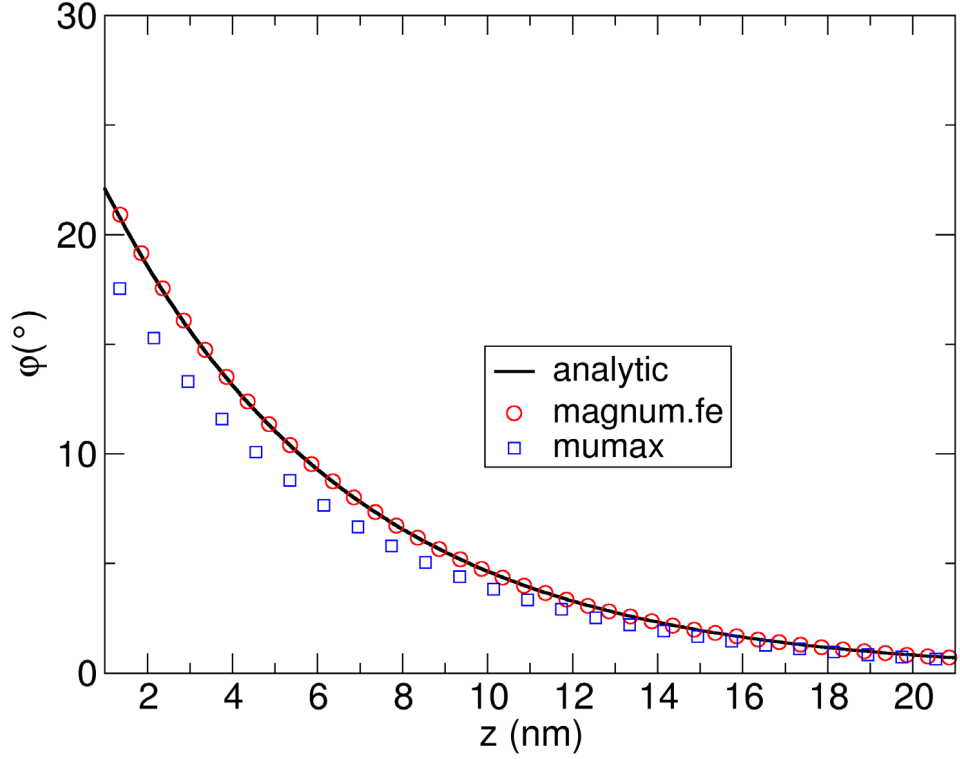

*Figure 4: Spin angle $\varphi$ as function of the z coordinate for $h_x = 4$ and a mesh size of $dz = 0.5$ nm for (magnum.fe) finite-element code and (mumax) a finite-difference code. The analytical solution is according to Eq.* (3.3) *and Eq.* (3.82).

The critical field $h_{x,crit}$, where the entire structure is perfectly saturated is derived in Appendix B as,

$$H_{x,crit} = \frac{2K_1}{J_s} h_{x,crit} = \frac{2K_1}{J_s} \lim_{\varphi_0 \to 0} \frac{1}{4} \frac{\cos(2\varphi_0) - 1 - \frac{J_{rkky}^2}{2AK_1} \sin^2(2\varphi_0)}{\cos(\varphi_0) - 1} = \frac{2K_1}{J_s}\left(1 + \frac{J_{rkky}^2}{AK_1}\right) \quad (3.5)$$

This critical field will be used to test various micromagnetic RKKY implementations.

## 4. Finite-difference discretization

In the following we will derive expressions for the exchange field at material interfaces. We start from the general expression for the exchange field that is given according to Eq. (3.56) by:

$$\mathbf{H}_{ex} = \frac{2}{J_s(\mathbf{x})} \nabla \cdot \left( A(\mathbf{x}) \nabla \mathbf{m} \right) \tag{3.6}$$

In the following we assume that in region 1 and region 2 different magnetic materials with different exchange constants are present, as shown in Figure 1. Within each region the exchange constant is assumed to be constant. Hence, within region 1 the following holds:

$$\mathbf{H}_{ex,1} = \frac{2}{J_s(\mathbf{x})} \nabla \cdot \left( A(\mathbf{x}) \nabla \mathbf{m} \right) = \frac{2A_1}{J_{s,1}} \Delta \mathbf{m} \tag{3.7}$$

In order to solve this problem we split the finite-difference solution in two parts. Part one for z<0 and part two for z>0. In the following we consider a 1d problem, where the magnetization only varies along the z-axis (e.g. a situation as shown in Figure 3) Eq. (3.7) simplifies to

$$\mathbf{H}_{ex,1}(z) = \frac{2A_1}{J_{s,1}} \mathbf{m}_1''(z) \tag{3.8}$$

For all discretization points which are not located at the boundary of the region a standard three-point stencil can be used for the evaluation of the second derivative m_1''(z).

The discretization point where the RKKY interaction comes into play is at, $z = -\Delta z / 2$ which we discuss in the following. To evaluate

$$\mathbf{H}_{ex}\left(-\frac{\Delta z}{2}\right) = \frac{2A_1}{J_{s,1}} \mathbf{m}_1''\left(-\frac{\Delta z}{2}\right) \tag{3.9}$$

we approximate the second derivative by the derivative of the first derivative,

$$\mathbf{m}_1''\left(-\frac{\Delta z}{2}\right) \approx \frac{\mathbf{m}_1'(0) - \mathbf{m}_1'(-\Delta z)}{\Delta z} \tag{3.10}$$

The first derivative with the domain at $z = -\Delta z$ can be simply approximate

$$\mathbf{m}_1'(-\Delta z) \approx \frac{\mathbf{m}_1\left(-\frac{\Delta z}{2}\right) - \mathbf{m}_1\left(-\frac{3\Delta z}{2}\right)}{\Delta z} \tag{3.11}$$

In order to calculate $\mathbf{m}_1'(0)$ at the interface, we introduce the unknown ghost point $\mathbf{m}_0 = \mathbf{m}_1(0)$ and write

$$\mathbf{m}_1'(0) \approx \frac{\mathbf{m}_0 - \mathbf{m}_1\left(-\frac{\Delta z}{2}\right)}{\Delta z / 2} \tag{3.12}$$

**Jump in the exchange constant $A$ – cell-based FD**

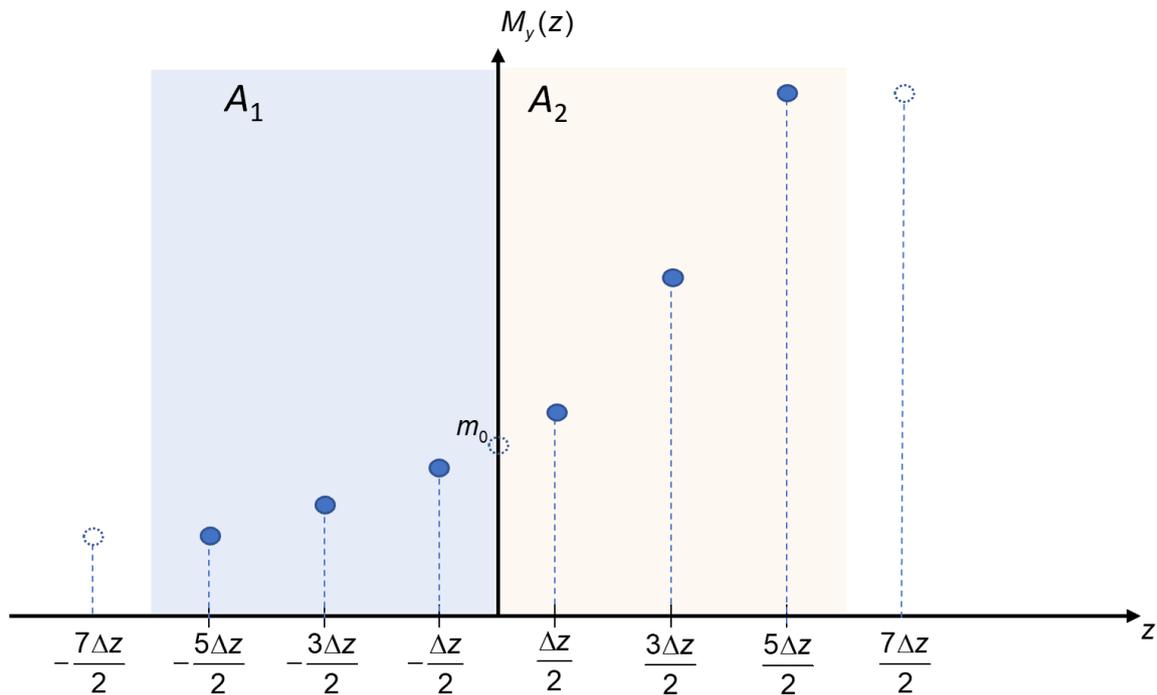

*Figure 5: Cell-based finite difference discretization when at the interface at z=0 the exchange constant exhibits a jump.*

All commonly used finite difference codes use a cell based approach [6]–[8], [12]. Special care has to be taken if jumps in the material parameters within the computed domain occur [9]. In the following we assume that the material parameter *A* jumps at z=0 as shown in Figure 5. Hence the magnetization is not differentiable at *z* = 0. The normal derivative of the magnetization shows a jump of

$$2A_1 (\nabla \mathbf{m}_1) \mathbf{n} = 2A_2 (\nabla \mathbf{m}_2) \mathbf{n} \tag{3.13}$$

Since,

$$\mathbf{m}_0 := \mathbf{m}_1(0) = \mathbf{m}_2(0) \tag{3.14}$$

We introduce the variable

$$\begin{aligned} \mathbf{m}(z) = \mathbf{m}_1(z) \quad &\text{for} \quad z \leq 0 \\ \mathbf{m}(z) = \mathbf{m}_2(z) \quad &\text{for} \quad z \geq 0 \end{aligned} \tag{3.15}$$

With the equations Eq. (3.13) and Eq. (3.14) we can calculate the unknown $\mathbf{m}_1(0)$ and $\mathbf{m}_2(0)$. From Eq. (3.13) it follows for the x component

$$A_1 \frac{\mathbf{m}_0 - \mathbf{m}\left(-\frac{\Delta z}{2}\right)}{\Delta z / 2} = A_2 \frac{\mathbf{m}\left(\frac{\Delta z}{2}\right) - \mathbf{m}_0}{\Delta z / 2} \tag{3.16}$$

From Eq. (3.16) $m_{z,0}$ can be calculated.

$$\mathbf{m}_0 = \frac{A_2 \mathbf{m}\left(\frac{\Delta z}{2}\right) + A_1 \mathbf{m}\left(-\frac{\Delta z}{2}\right)}{A_1 + A_2} \tag{3.17}$$

Substituting $m_0$ into Eq. (3.12) and Eq, (3.9) and neglecting all terms linear in m(-Delta z/2) one obtains

$$\mathbf{H}_{ex}\left(-\frac{\Delta z}{2}\right) = \frac{2A_1}{J_{s,1}\Delta z^2} \left[ \frac{2A_1 \mathbf{m}(-3\Delta z / 2)}{A_1 + A_1} + \frac{2A_2 \mathbf{m}(\Delta z / 2)}{A_1 + A_2} \right] \tag{3.18}$$

(3.18) $m_x(-\Delta z/2)$ Neglecting terms linear in $\mathbf{m}(-\Delta z/2)$ in the effective field is justified by the fact that they do not change the dynamics of the Landau – Lifshitz – Gilbert equation at $\frac{d}{dt}\mathbf{m}(-\Delta z/2)$.

It is worth noting that Eq. (3.18) can be used everywhere in space. If in some region $A := A_1 = A_2$ it reduces to the well known equation for the exchange field.

$$\mathbf{H}_{ex}\left(k\frac{\Delta z}{2}\right) = \frac{2A}{J_s \Delta z^2}\left[\mathbf{m}\left((k-1)\Delta z/2\right) + \mathbf{m}\left((k+1)\Delta z/2\right)\right] \qquad (3.19)$$

**Coupling via RKKY interface – cell-based FD**

Besides jump in the material parameters another practical important case is the inclusion of RKKY interaction, which might leads to antiparallel or weak coupling between these coupled layers [4], [13]. In the following we will derive a zero, first and second order method to treat these interactions.

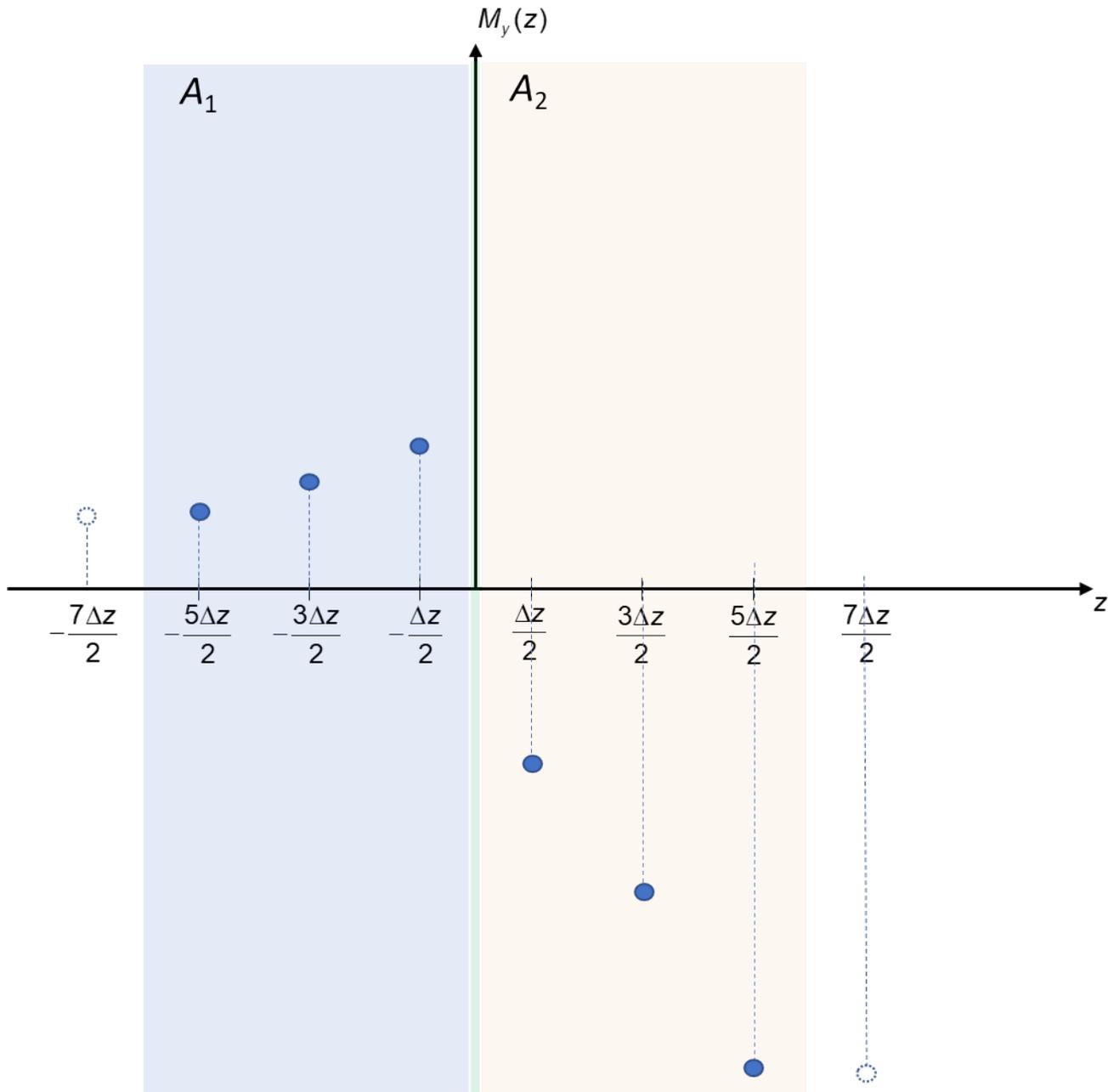

*Figure 6: Finite-difference discretization for cell-based finite-difference method. The discretization points for the magnetization are loacted at the cell centers. The spacer layer which couples the magnetization **m** (-Δz/2) and the magnetization **m**(Δz/2) is assumed to be infinitelz thin. The thickness of the two magnets is in this example 3Δz each.*

We assume an infinitely thin spacer layer at *z* = 0 as shown in Figure 6. The exchange field is given by

$$\mathbf{H}_{ex}\left(-\frac{\Delta z}{2}\right) = \frac{2A}{J_s}\mathbf{m}_1''\left(-\frac{\Delta z}{2}\right) \quad (3.20)$$

Where we approximate

$$\mathbf{m}_1''\left(-\frac{\Delta z}{2}\right) = \frac{\mathbf{m}_1'(0) - \mathbf{m}_1'(-\Delta z)}{\Delta z} \quad (3.21)$$

The boundary condition at z = 0 is given by,

$$\mathbf{m}_1'(0) = -\frac{J_{rkky}}{2A_1}\left[\mathbf{m}_1(0)\mathbf{m}_2(0)\right]\mathbf{m}_1(0) + \frac{J_{rkky}}{2A_1}\mathbf{m}_2(0) \quad (3.22)$$

Here it is important to note, that it is important to consider here the first term on the right hand side of Eq. (3.22). If this term is neglected significantly deviation of the correct interpolation are obtained. Only for the interpolation order O(0) this term cancels out and hence has not to be considered.

One obtains:

$$\mathbf{H}_{ex}\left(-\frac{\Delta z}{2}\right) = \frac{J_{rkky}\left[\mathbf{m}_2(0) - \left[\mathbf{m}_1(0)\mathbf{m}_2(0)\right]\mathbf{m}_1(0)\right] - 2A\mathbf{m}_1'(-\Delta z)}{J_s \Delta z} \quad (3.23)$$

The second term on the right side of Eq. (3.21) it is simply obtained by a finite difference approximation

$$\mathbf{m}_1'(-\Delta z) = \frac{\mathbf{m}_1\left(-\frac{\Delta z}{2}\right) - \mathbf{m}_1\left(-\frac{3\Delta z}{2}\right)}{\Delta z} \quad (3.24)$$

For the evaluation of $\mathbf{m}_1(0)$ and $\mathbf{m}_2(0)$ different interpolation orders can be used:

**Order (0):** The magnetization is simply approximated as:

$$\mathbf{m}_2(0) \approx \mathbf{m}_2\left(\frac{\Delta z}{2}\right) \tag{3.25}$$

By neglecting all terms parallel to $\mathbf{m}_1\left(-\frac{\Delta z}{2}\right)$ (hence the first term on the right hand side of Eq. (3.22) cancels out, which is a commonly used interpolation in O(0) codes )we get,

$$\mathbf{H}_{ex}\left(-\frac{\Delta z}{2}\right) = \frac{1}{J_s \Delta z^2}\left[J_{rkky}\Delta z \mathbf{m}_2\left(\frac{\Delta z}{2}\right) + 2A_1\mathbf{m}_1\left(-\frac{3\Delta z}{2}\right)\right] \tag{3.26}$$

**Order (1):** We approximate the magnetization in region 1 and region 2 with a series expansion first order using forward (backwards) difference approximation[14]:

$$\mathbf{m}_1(0) \approx \mathbf{m}_1\left(-\frac{\Delta z}{2}\right) + \frac{\Delta z}{2}\mathbf{m}_1{}'\left(-\frac{\Delta z}{2}\right) \approx \mathbf{m}_1\left(-\frac{\Delta z}{2}\right) + \frac{\Delta z}{2}\left(\frac{\mathbf{m}_1\left(-\frac{\Delta z}{2}\right) - \mathbf{m}_1\left(-\frac{3\Delta z}{2}\right)}{\Delta z}\right) \tag{3.27}$$

$$\mathbf{m}_2(0) \approx \mathbf{m}_2\left(\frac{\Delta z}{2}\right) - \frac{\Delta z}{2}\mathbf{m}_2{}'\left(\frac{\Delta z}{2}\right) \approx \mathbf{m}_2\left(\frac{\Delta z}{2}\right) - \frac{\Delta z}{2}\left(\frac{\mathbf{m}_2\left(\frac{3\Delta z}{2}\right) - \mathbf{m}_2\left(\frac{\Delta z}{2}\right)}{\Delta z}\right) \tag{3.28}$$

This interpolation is used within Eq. (3.23) for the calculation of the effective field at the boundary cells.

**Order (2):** Increasing the interpolation order to second order using forward difference approximation of $\mathbf{m}_1{}'(-\Delta z/2)$ leads to,

$$\begin{aligned}\mathbf{m}_1(0) &\approx \mathbf{m}_1\left(-\frac{\Delta z}{2}\right) + \frac{\Delta z}{2}\mathbf{m}_1{}'\left(-\frac{\Delta z}{2}\right) \\ &\approx \mathbf{m}_1\left(-\frac{\Delta z}{2}\right) + \frac{\Delta z}{2}\left(\frac{3\mathbf{m}_1\left(-\frac{\Delta z}{2}\right) - 4\mathbf{m}_1\left(-\frac{3\Delta z}{2}\right) + \mathbf{m}_1\left(-\frac{5\Delta z}{2}\right)}{2\Delta z}\right)\end{aligned} \tag{3.29}$$

Equivalently one obtains using backward difference approximation the approximation for $m_2(0)$:

$$\mathbf{m}_2(0) \approx \mathbf{m}_2\left(\frac{\Delta z}{2}\right) - \frac{\Delta z}{2}\mathbf{m}_2{}'\left(\frac{\Delta z}{2}\right)$$

$$\approx \mathbf{m}_2\left(\frac{\Delta z}{2}\right) - \frac{\Delta z}{2}\left(\frac{-3\mathbf{m}_2\left(\frac{\Delta z}{2}\right) + 4\mathbf{m}_2\left(\frac{3\Delta z}{2}\right) - \mathbf{m}_2\left(\frac{5\Delta z}{2}\right)}{2\Delta z}\right) \quad (3.30)$$

It is worth noting that for the cell-based finite-difference method the ghost-cell has to be set on the left boundary as follows (an equivalent condition is used for the right boundary)

$$\mathbf{m}_{-7\Delta z/2} = \mathbf{m}_{-5\Delta z/2} \quad (3.31)$$

If

$$\mathbf{m}_{-7\Delta z/2} = \mathbf{m}_{-3\Delta z/2} \quad (3.32)$$

is used as boundary condition a significant deviation of the results is obtained as shown in Figure 8 (cell, O1, low order BC).

**Coupling via RKKY interface – node based FD**

Since the inclusion of RKKY interaction is more naturally possible within a node based finite difference scheme, we also show this approach here.

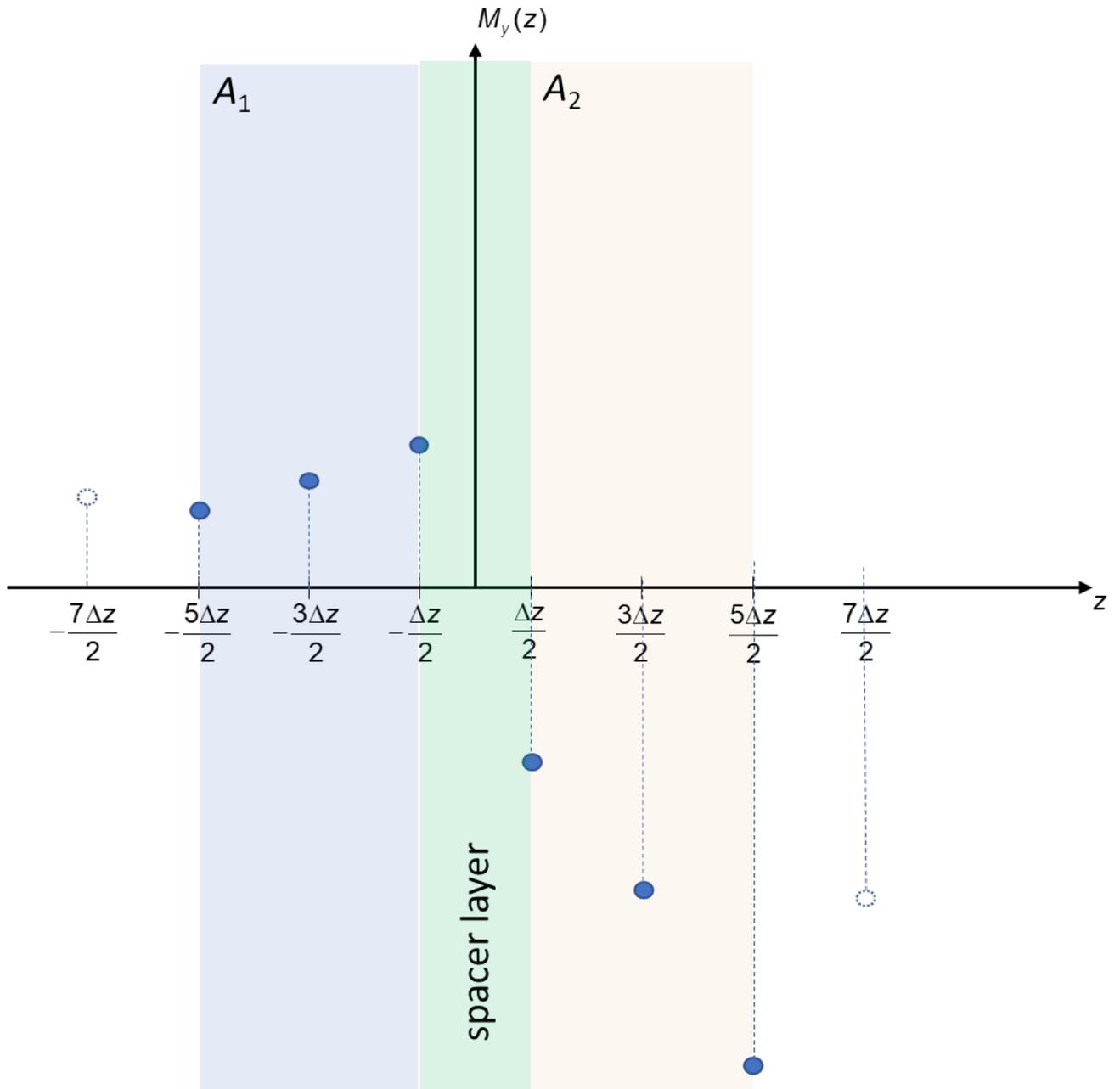

*Figure 7: Finite-difference discretization for node based finite-difference method. The magnetization is given on the mesh nodes. The spacer layer which couples the magnetization **m** (-Δz/2) and the magnetization **m**(Δz/2) has a thickness of Δz. The thickness of the two magnets is in this example 2Δz each.*

We assume a spacer layer between the nodes at $z = -\Delta z/2$ and $z = \Delta z/2$ as shown in . The coupling between the nodes is given by Eq. (2.4) which leads to the boundary condition given by Eq. (2.8). In order to derive the exchange field at $z = -\Delta z/2$ we again start from

$$\mathbf{H}_{ex}\left(-\frac{\Delta z}{2}\right) = \frac{2A}{J_s}\mathbf{m}_1''\left(-\frac{\Delta z}{2}\right) \tag{3.33}$$

Where we approximate

$$\mathbf{m}_1''\left(-\frac{\Delta z}{2}\right) = \frac{\mathbf{m}_1'\left(-\frac{\Delta z}{2}\right) - \mathbf{m}_1'(-\Delta z)}{\Delta z / 2} \tag{3.34}$$

From the boundary condition Eq. (2.8) at $z = -\Delta z / 2$ follows by neglecting again all terms parallel to $\mathbf{m}_1\left(-\frac{\Delta z}{2}\right)$ (since it does not change the dynamics of the system),

$$\mathbf{m}_1'\left(-\frac{\Delta z}{2}\right) = \frac{J_{rkky}}{2A_1}\mathbf{m}_2\left(\frac{\Delta z}{2}\right) \tag{3.35}$$

Hence,

$$\mathbf{H}_{ex}\left(-\frac{\Delta z}{2}\right) = \frac{2}{J_s \Delta z}\frac{J_{rkky}\mathbf{m}_2\left(\frac{\Delta z}{2}\right) - 2A\mathbf{m}_1'(-\Delta z)}{} \tag{3.36}$$

The second term on the right side of Eq. is simply obtained by a finite difference approximation

$$\mathbf{m}_1'(-\Delta z) = \frac{\mathbf{m}_1\left(-\frac{\Delta z}{2}\right) - \mathbf{m}_1\left(-\frac{3\Delta z}{2}\right)}{\Delta z} \tag{3.37}$$

Hence, we obtain for the exchange field

$$\mathbf{H}_{ex}\left(-\frac{\Delta z}{2}\right) = \frac{2}{J_s \Delta z^2}\left(J_{rkky}\Delta z\, \mathbf{m}\left(\frac{\Delta z}{2}\right) + 2A_1 \mathbf{m}\left(-\frac{3\Delta z}{2}\right)\right) \tag{3.38}$$

It is worth noting that this formula for the RKKY interface contains a term 4A, whereas the formula within the bulk of the magnet contains a term of 2A as it can be seen in Eq. (3.19). In Appendix C it is shown that in 1D systems the calculation of the effective field according to Eq. is equivalent to an 1D finite element implementation of RKKY interaction.

For this nodal based finite-difference method, the ghost cell has to be set on the left boundary as follows (an equivalent condition is used for the right boundary)

$$\mathbf{m}_{-7\Delta z/2} = \mathbf{m}_{-3\Delta z/2} \quad (3.39)$$

If

$$\mathbf{m}_{-7\Delta z/2} = \mathbf{m}_{-5\Delta z/2} \quad (3.40)$$

is used as boundary condition a significant deviation of the results is obtained as shown in Figure 8 (node, low order BC).

### 5. *Comparison of analytical solution with micromagnetic simulations*

#### a. Limit for thin layers – homogenous magnetization within the layers

All tested micromagnetic codes were able to accurately reproduce the hysteresis loop of two antiferromagnetic thin layers as shown in Figure 8 (material parameters are listed in caption of Figure 8). This test case was primarily chosen to validate the different definitions of the exchange implementations of the codes, that correspond to:

$$E_{rkky,OOMMF}(\mathbf{m}_1, \mathbf{m}_2) = -2 \int_{\partial\Omega_1 \cap \partial\Omega_2} J_{rkky} \mathbf{m}_1 \mathbf{m}_2 dA \quad (3.41)$$

$$E_{rkky,\text{mumax}}(\mathbf{m}_1, \mathbf{m}_2) = - \int_{\partial\Omega_1 \cap \partial\Omega_2} J_{rkky} \mathbf{m}_1 \mathbf{m}_2 dA \quad (3.42)$$

However, it is interesting to note that the proper implementation of the Neuman boundary condition at the air surface is important for accurate results. Figure 8 (paper, node) shows the node based RKKY implementation according to Eq. and the boundary condition according to Eq. . Here perfect agreement with the analytic result (2.15) is observed. In contrast Figure 8 (paper, node, low order BC) shows the same node-based RKKY implementation according to Eq. but a different boundary condition according to Eq. . Here, a significant error occurs. An equivalent situation can be observed with the cell-based implementation.

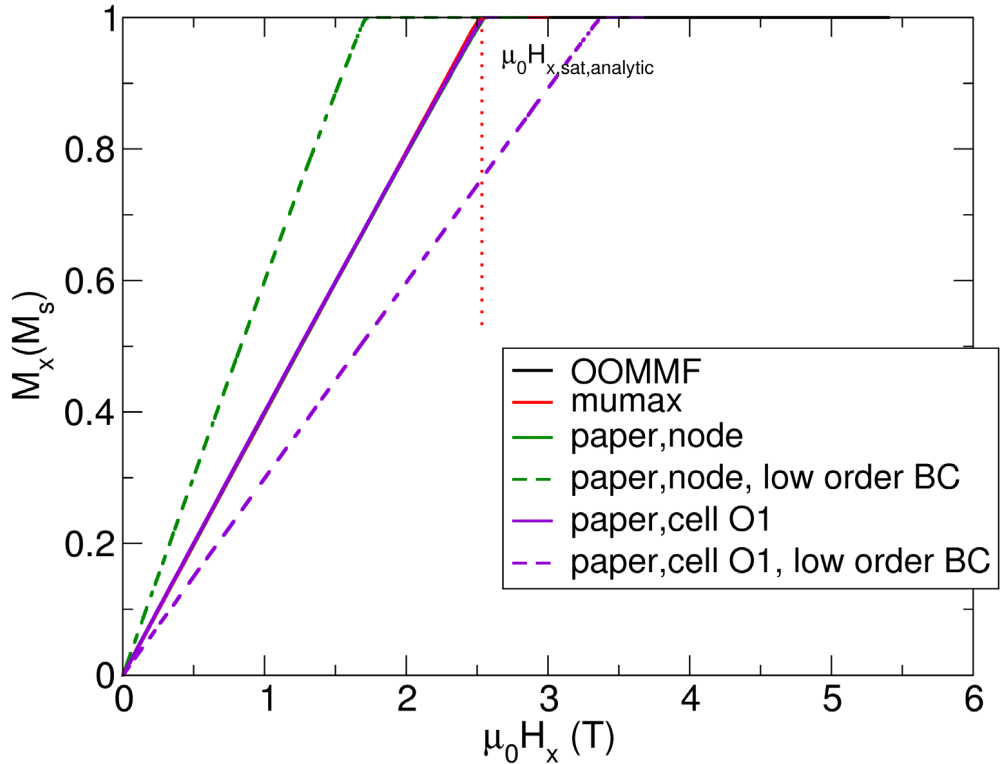

*Figure 8: Hysteresis loop of two antiferromagnetic coupled thin films, where each film has a thickness of t = 2nm, A=5e-11 J/m, K1 = 0, Js = 1T. The analytic result is obtained from Eq. (2.15). Different finite-difference implementations are tested. (OOMMF) $J_{rkky,OOMMF}$ = -0.001 J/m (mumax) solution $J_{rkky,OOMMF}$ = -0.002 J/m (paper, cell O1) cell-based RKKY implementation according to Eq. and boundary condition according to Eq. (3.31) and $J_{rkky,this\,paper}$ = -0.002 J/m. (paper, cell O1, low order BC) cell based RKKY implementation according to Eq. and boundary condition according to (3.32). (paper, node) node based RKKY implementation according to Eq. and the boundary condition according to Eq. . (paper, node, low order BC) node based RKKY implementation according to Eq. and the boundary condition according to Eq. .*

b. Limit for thick layers – inhomogenous magnetization within the layers

The test cases for thin magnetic layers, where no partial domain walls in the perpendicular direction arise within the magnetic layers can be well reproduced by all micromagnetic codes. The situation is different if inhomogeneous states are formed within the magnetic layers. To realize this situation we use the example where we derived the analytic solution for the saturation field according to Eq. (3.5).

In the following, we perform micromagnetic simulations for the parameters $A = 10^{-11}$ J/m, $K_1 = 10^5$ J/m³, $J_s = 1$T and $J_{rkky} = -2\times10^{-3}$ J/m using the finite-element code magnum.fe [15] and different state of the art finite-difference codes (mumax and OOMMF) as well as different RKKY implementations according to this paper. The external field is slowly increased within 20 ns from $h_x = 4$ to $h_x = 6$, where $h_x$ is the field in units of $2K_1 / J_s$. The hysteresis loop close to the saturation field is shown in Figure 9. Excellent agreement between magnum.fe and the analytic solution, which is $h_{x,crit} = 5$ can be obtained for various mesh sizes as shown in Figure 10.

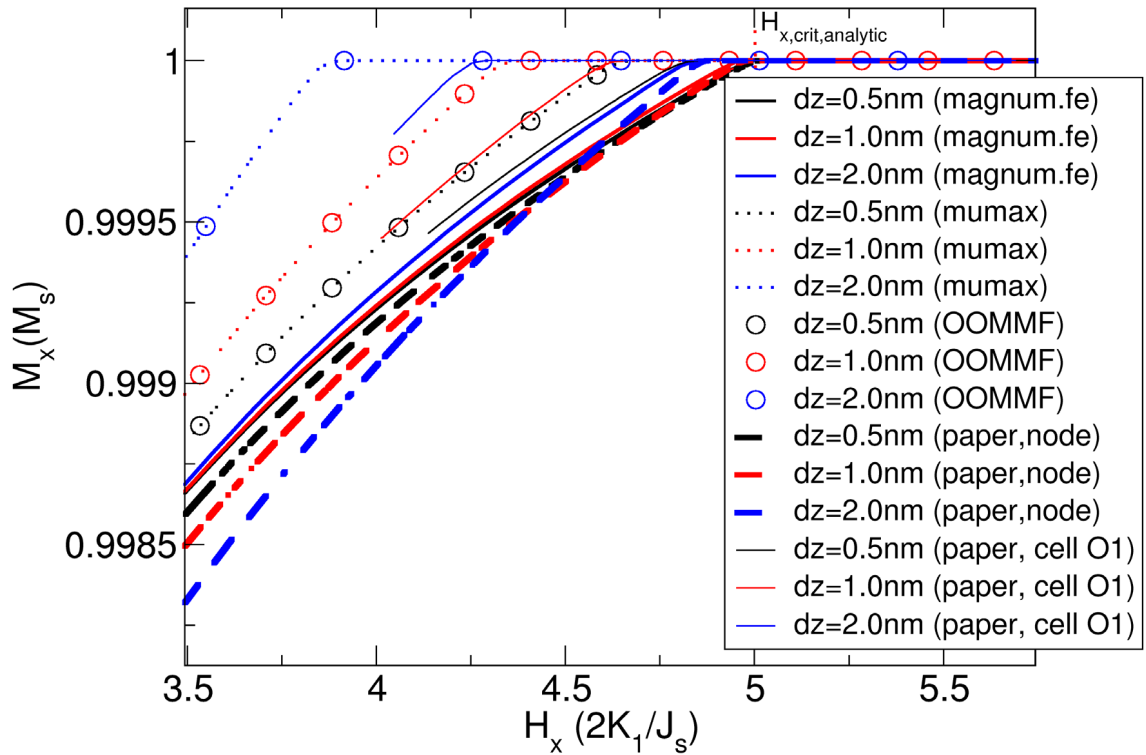

Figure 9: (a) Micromagnetically obtained Mx(Hx) loop for a structure with lateral dimensions -400<z<400, -1<x<1, -<y<1 for different mesh sizes and different codes. (magnum.fe) is a finite-element code with linear shape functions for the magnetization between the node points (mumax) finite-difference code with cell-centered magnetization (OOMMF) finite-difference code with the same interpolation as mumax (paper,node) nodal based finite-difference code according to Eq. for the RKKY interaction (paper,cell O1) cell-based finite-difference code according to *Eq.* for the RKKY interaction. The analytically obtained saturation field is $H_{x,crit,analytic}$ = 5 x $2K_1/J_s$.

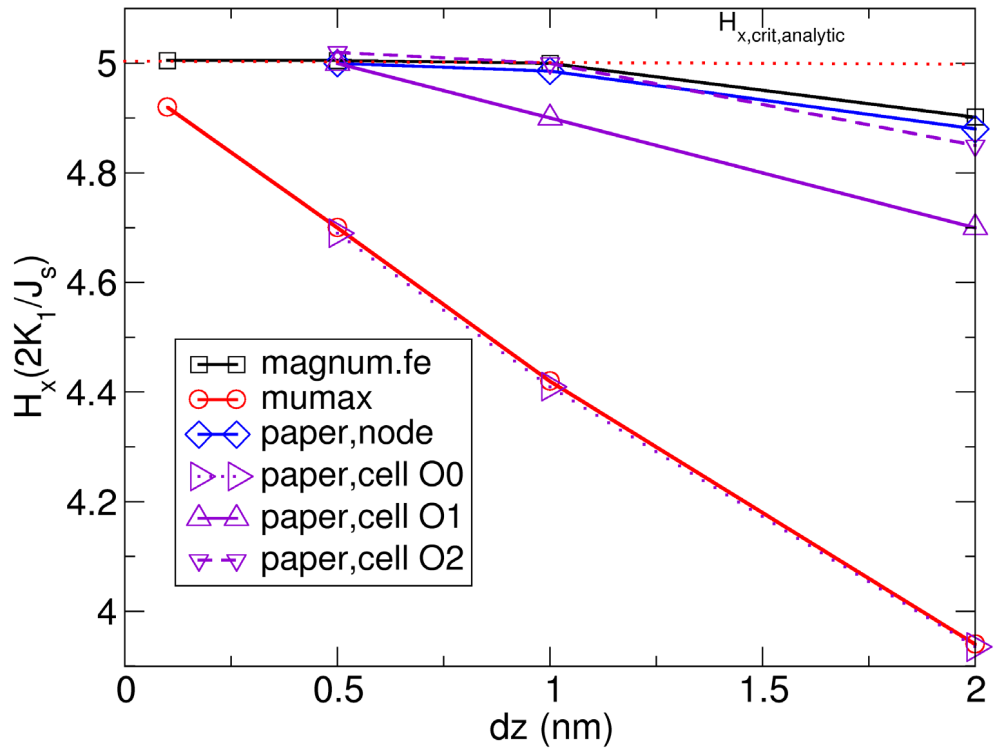

*Figure 10: Obtained saturation field for the same example as Figure 9 for different discretization lengths for different micromagnetic methods. (mumax) Only in the limit of infinitely small mesh size a good agreement with the analytic value is obtained (magnum.fe) the finite element method shows a much better convergence (paper.node) The finite difference implementation according to Eq. , which is equivalent to the finite element 1D case shows similarly good convergence.* (paper,cell O0) the cell centered finite-difference method order zero of the interpolation of $m_0$ according to Eq. (3.26) (paper,cell O1) the cell centered finite difference method of order one according *Eq.* (paper,cell O2) the cell centered finite difference method order two according to Eq (3.30).

Also the (paper, node) implementation shows a good convergence towards to the analytic solution. Since we show in Appendix C that the (paper, node) implementation is equivalent to an 1D finite element implementation the similar behaviour to the 3d finite-element code magnum.fe is not surprising.

Interestingly the mumax result, which is equivalent to the OOMMF result show that the analytic solution in only obtained for infinitely small mesh sizes. For a discretization length of 2 nm which is already a fine mesh size for common micromagnetic simulations, a significant error is obtained. The exchange length which is usually a guide for the required discretization length amounts to $l_{ex} = \sqrt{A/K_1} = 10 nm$ in this example. Hence, the discretization length of 2 nm is five times smaller than the exchange lengths still leads to an error in the saturation field of more than 20% in standard FD codes. It is worth noting that this error also occurs for small magnetization inhomogeneities formed as shown in Figure 11.

This large error is reduced successfully for the proposed first order and even better for the second order cell based finite difference codes. It is important to note that in this proposed implementation, the first term on the right hand side of Eq. (3.22) has to be considered, which contains $m_1(0)$ and $m_2(0)$ in the Robin boundary condition, which is generally in state of the art O(0) implementation neglected.

## 6. Conclusion

In this paper we investigate different implementations of RKKY interaction in finite-difference micromagnetic codes. While all codes deliver reliable results for the case that the RKKY coupling acts between thin ferromagnetic layers, significant errors occur if partial domain walls are formed in the coupled ferromagnetic layers. Partial domain walls are commonly formed in ferrimagnetic materials that are coupled antiferromagnetically to ferromagnetic layers such as in systems of Ref. [5]. Finite-element codes show good convergence to the correct solutions. Similarly, nodal based finite difference codes can properly describe these antiferromagnetically coupled thick layers. If cell-based finite difference codes are used the most commonly used codes exhibit significant errors. The reason is that the boundary condition due to RKKY interaction is not imposed at the surface of the magnet, where it has to be imposed but on the cell centre of the finite difference cell at the boundary. We show by using first and second order interpolation how the boundary condition can be correctly imposed at the surface of the domain. The second order cell based scheme gives similar results as the first order nodal based scheme.

The financial support of the FWF – projects P 34671 and I 4917 is acknowledged.

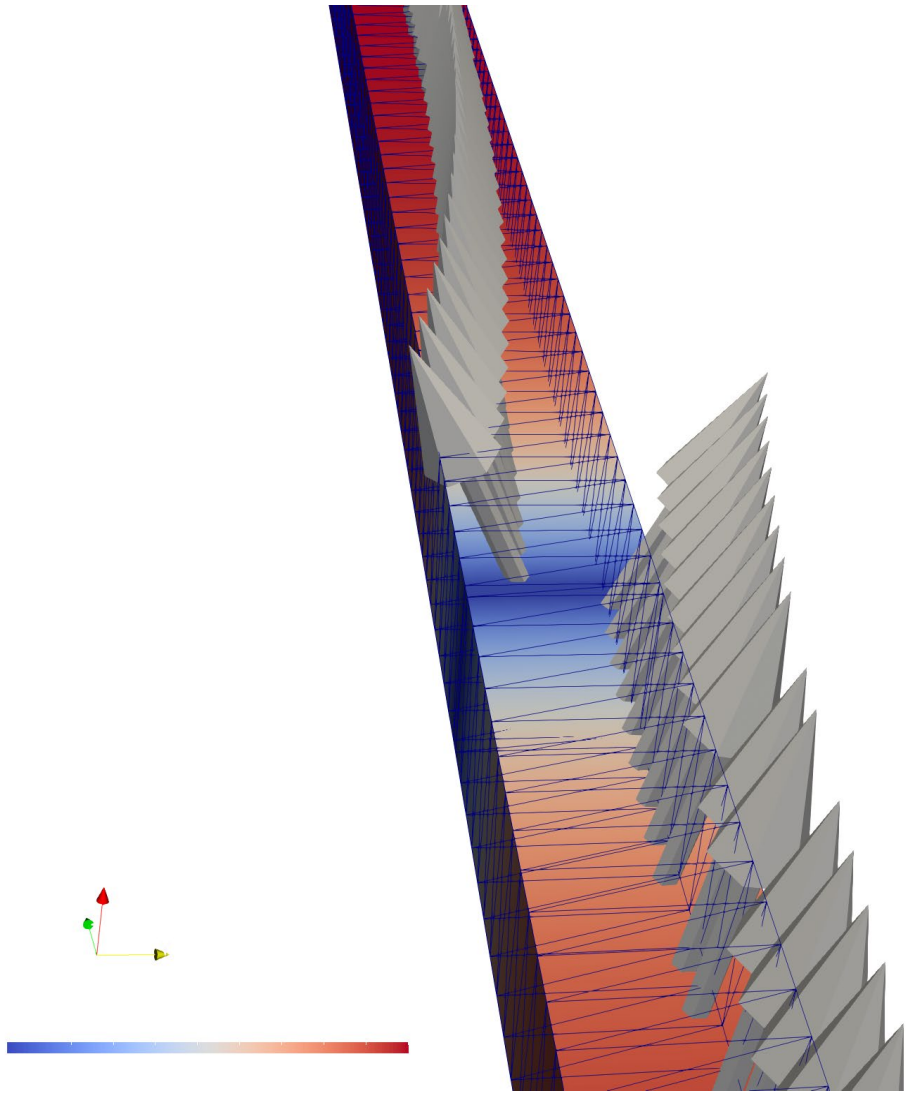

Figure 11: magnum.fe solution for $h_x$ = 3.9 and dz = 0.5 nm. For this field value the mumax solution for dz = 2.0 nm the magnetization points already everywhere in $M_z$=1 direction.

# Appendix A: Variation of the total energy to obtain the equilibrium miromagnetic equation

In order to obtain the equilibrium equation for the magnetization at the surface and volume of the magnet we calculate the variation of the total energy $E_{tot}$ given by Eq. (2.5) with the constraint of the norm of the magnetization

$$|\mathbf{m}_i| = 1 \tag{3.43}$$

The constraint of the magnetization is included in the variation of the total energy by introducing Lagrange multiplier $\lambda_1$ and $\lambda_2$ within the volumes and $\lambda_{s,1}$ and $\lambda_{s,2}$ at the surfaces. The energy functional with the Lagrange multipliers is,

$$\begin{aligned}
\tilde{E}_{tot}\left(\mathbf{m}_1, \mathbf{m}_2, \lambda_1, \lambda_{s,1}, \lambda_2, \lambda_{s,2}\right) &= E_{tot}\left(\mathbf{m}_1, \mathbf{m}_2\right) \\
&+ \int_{\Omega_1} \lambda_1 \left(1 - m_{x,1}^2 - m_{y,1}^2 - m_{z,1}^2\right) dV + \int_{\partial\Omega_1} \lambda_{s,1}\left(1 - m_{x,1}^2 - m_{y,1}^2 - m_{z,1}^2\right) dA + \\
&+ \int_{\Omega_2} \lambda_2 \left(1 - m_{x,2}^2 - m_{y,2}^2 - m_{z,2}^2\right) dV + \int_{\partial\Omega_2} \lambda_{s,2}\left(1 - m_{x,2}^2 - m_{y,2}^2 - m_{z,2}^2\right) dA
\end{aligned} \tag{3.44}$$

Here, the magnetization vector within the volume $\Omega_1$ (accordingly for $\Omega_2$) is

$$\mathbf{m}_1 = \begin{pmatrix} m_{x,1} \\ m_{y,1} \\ m_{z,1} \end{pmatrix} \tag{3.45}$$

For the variation of the x- component of the normalized magnetization $m_1$ we get,

$$\begin{aligned}
\tilde{E}_{tot}\left(m_{x,1} + \delta m_{x,1}, m_{y,1}, m_{z,1}, \mathbf{m}_2, \lambda_1, \lambda_{s,1}, \lambda_2, \lambda_{s,2}\right) &= E_{tot}\left(m_{x,1} + \delta m_{x,1}, m_{y,1}, m_{z,1}, \mathbf{m}_2\right) \\
&+ \int_{\Omega_1} 2\lambda_1 m_{x,1} \delta m_{x,1} dV + \int_{\partial\Omega_1} 2\lambda_{s,1} m_{x,1} \delta m_{x,1} dA + non\ linear\ tearms\ in\ \delta m_{x,1}
\end{aligned} \tag{3.46}$$

For the variation of the term $E_{tot}\left(m_{x,1} + \delta m_{x,1}, m_{y,1}, m_{z,1}, \mathbf{m}_2\right)$ we start with the exchange energy :

$$\begin{aligned}
E_{ex} &= \int_{\Omega} A_{x,1}\left[\left(\nabla m_{x,1} + \nabla \delta m_{x,1}\right)\left(\nabla m_{x,1} + \nabla \delta m_{x,1}\right)\right] dV \\
&= \int_{\Omega_1} A_{x,1}\left[\left(\nabla m_{x,1}\right)^2 + 2\nabla m_{x,1} \nabla \delta m_{x,1} + \left(\nabla \delta m_{x,1}\right)^2\right] dV
\end{aligned} \tag{3.47}$$

Applying Green's first identity to transform the term $2\nabla m_{x,1} \nabla \delta m_{x,1}$ to a linear term in $\delta m_{x,1}$ and keeping only the linear terms in $\delta m_{x,1}$ we obtain

$$E_{ex} = -\int_{\Omega_1} 2\nabla \cdot \left(A_{x,1} \nabla m_{x,1}\right) \delta m_{x,1} dV + \int_{\partial \Omega_1} A_{x,1} \left[2\nabla m_{x,1} \delta m_{x,1}\right] \mathbf{n} dA \tag{3.48}$$

For the total energy we consider here all terms that leads to surface contributions, which are the exchange energy and the RKKY interaction. If we consider only the linear terms in $\delta m_{x,1}$ we obtain

$$\begin{aligned}
\tilde{E}_{tot}&\left(m_{x,1} + \delta m_{x,1}, m_{y,1}, m_{z,1}, \mathbf{m}_2, \lambda_1, \lambda_{s,1}, \lambda_2, \lambda_{s,2}\right) = \\
&-\int_{\Omega_1} 2\nabla \cdot \left(A_{x,1} \nabla m_{x,1}\right) \delta m_{x,1} dV + \int_{\partial \Omega_1} A_{x,1} \left[2\nabla m_{x,1} \delta m_{x,1}\right] \mathbf{n} dA \\
&- \int_{\partial \Omega_1 \cap \partial \Omega_2} J_{rkky} \delta m_{x,1} \, m_{2,x} dA \\
&+ \int_{\Omega_1} 2\lambda_1 m_{x,1} \delta m_{x,1} dV + \int_{\partial \Omega_1} 2\lambda_{s,1} m_{x,1} \delta m_{x,1} dA
\end{aligned} \tag{3.49}$$

From Eq. (3.49) we get within the volume $\Omega_1$ of the magnet:

$$2\lambda_1 m_{x,1} - 2\nabla \cdot \left(A_{x,1} \nabla m_{x,1}\right) = 0 \tag{3.50}$$

$$2\lambda_1 m_{y,1} - 2\nabla \cdot \left(A_{y,1} \nabla m_{y,1}\right) = 0 \tag{3.51}$$

$$2\lambda_1 m_{z,1} - 2\nabla \cdot \left(A_{z,1} \nabla m_{z,1}\right) = 0 \tag{3.52}$$

$$m_{x,1}^2 + m_{y,1}^2 + m_{z,1}^2 = 1$$

Eq. (3.50) to Eq. (3.52) can be written in the form

$$\mathbf{m}_1 \times 2\nabla \cdot \left(\mathbf{A}_1 \nabla \mathbf{m}_1\right) = 0 \tag{3.53}$$

where

$$\mathbf{A}_1 = \begin{pmatrix} A_{x,1} & 0 & 0 \\ 0 & A_{y,1} & 0 \\ 0 & 0 & A_{z,1} \end{pmatrix} \tag{3.54}$$

The other energy contributing terms that do not contain derivates in the energy are trivially included and only lead to contributions to the volume equilibrium condition. One obtains for the volume $\Omega_i$,

$$\mathbf{m}_i \times \left[ 2\nabla \cdot (\mathbf{A}_i \nabla \mathbf{m}_i) + J_{s,i} \mathbf{H}_{ext} + 2K_{1,1} (\mathbf{m}_i \mathbf{k}_i) \mathbf{k}_i \right] = 0 \qquad (3.55)$$

Which is convenient to write

$$\mathbf{m}_i \times \underbrace{\left[ \frac{2\nabla \cdot (\mathbf{A}_i \nabla \mathbf{m}_i)}{J_{s,i}} + \mathbf{H}_{ext} + \frac{2K_{1,i}}{J_{s,i}} (\mathbf{m}_i \mathbf{k}_i) \mathbf{k}_i \right]}_{\mathbf{H}_{eff}} = 0 \qquad (3.56)$$

**Surface**

From Eq. (3.49) we get on the surface of the magnet $\partial\Omega_1$

$$E_{tot} = \int_{\partial\Omega_1} A_{1,x} \left[ 2\nabla m_{x,1} \delta m_{x,1} \right] \mathbf{n} dA - \int_{\partial\Omega_1 \cap \partial\Omega_2} J_{rkky} \delta m_{x,1} m_{x,2} dA + \int_{\partial\Omega_1} 2\lambda_{s,1} m_{x,1} \delta m_{x,1} dA \qquad (3.57)$$

Hence, we get the boundary condition at $\partial\Omega_1 \cap \partial\Omega_2$:

$$2\mathbf{A}_1 (\nabla \mathbf{m}_1) \mathbf{n} - J_{rkky} \mathbf{m}_2 = 2\lambda_{s,1} \mathbf{m}_1 \qquad (3.58)$$

Which leads to

$$\mathbf{m}_1 \times \left[ 2\mathbf{A}_1 (\nabla \mathbf{m}_1) \mathbf{n} - J_{rkky} \mathbf{m}_2 \right] = 0 \qquad (3.59)$$

and for the boundary $\partial\Omega_1 \setminus (\partial\Omega_1 \cap \partial\Omega_2)$:

$$\mathbf{m}_1 \times \left[ 2\mathbf{A}_1 (\nabla \mathbf{m}_1) \mathbf{n} \right] = 0 \qquad (3.60)$$

We can rewrite (3.60) by building the cross product with *m*₁:

$$\mathbf{m}_1 \times \left[ \mathbf{m}_1 \times \left[ 2\mathbf{A}_1 (\nabla \mathbf{m}_1) \mathbf{n} - J_{rkky} \mathbf{m}_2 \right] \right] = 0 \qquad (3.61)$$

$$\underbrace{\left[\underbrace{2\mathbf{A}_1\mathbf{m}_1(\nabla\mathbf{m}_1)\mathbf{n}}_{n\mathbf{A}_1\nabla\mathbf{m}_1^2}\right]}_{0}\mathbf{m}_1 - \underbrace{(\mathbf{m}_1\mathbf{m}_1)}_{1}2\mathbf{A}_1(\nabla\mathbf{m}_1)\mathbf{n} - (\mathbf{m}_1 J_{rkky}\mathbf{m}_2)\mathbf{m}_1 + \underbrace{(\mathbf{m}_1\mathbf{m}_1)}_{1}J_{rkky}\mathbf{m}_2 = 0 \quad (3.62)$$

Hence, one obtains the following Robin boundary conditions at the common surface $\partial\Omega_1 \cap \partial\Omega_2$:

$$2\mathbf{A}_1(\nabla\mathbf{m}_1)\mathbf{n} = -(\mathbf{m}_1 J_{rkky}\mathbf{m}_2)\mathbf{m}_1 + J_{rkky}\mathbf{m}_2 \quad (3.63)$$

## Appendix B: Derivation of analytic solution of RKKY coupled 1D wire

In order to obtain the analytical solution for the 1d wire of chapter 3 (Figure 1) we start from the effective field within the volume.

$$\mathbf{H}_{eff} = \frac{2A}{J_s}\begin{pmatrix}\frac{\partial^2}{\partial^2 z}\cos(\varphi(z)) \\ \frac{\partial^2}{\partial^2 z}\sin(\varphi(z)) \\ 0\end{pmatrix} + \begin{pmatrix}H_x \\ 0 \\ 0\end{pmatrix} + \frac{2K_1}{J_s}\begin{pmatrix}0 \\ \sin(\varphi(z)) \\ 0\end{pmatrix} \quad (3.64)$$

Using the notation $\varphi = \varphi(z)$ and using the equilibrium condition

$$\mathbf{m}\times\mathbf{H}_{eff} = \begin{pmatrix}\cos(\varphi) \\ \sin(\varphi) \\ 0\end{pmatrix}\times\mathbf{H}_{eff} \quad (3.65)$$

we see that all component of Eq. (3.65) are zero except the *z*-component for which we obtain,

$$\frac{2A}{J_s}\frac{\partial^2\varphi}{\partial^2 z} - H_x\sin(\varphi) + \frac{K_1}{J_s}\sin(2\varphi) = 0 \quad (3.66)$$

We multiply Eq. (3.66) with $\frac{\partial\varphi}{\partial z}$ and obtain

$$\frac{2A}{J_s}\underbrace{\frac{\partial^2\varphi}{\partial^2 z}\frac{\partial\varphi}{\partial z}}_{\frac{1}{2}\frac{\partial}{\partial z}\left[\frac{\partial\varphi}{\partial z}\right]^2} - H_x\underbrace{\sin(\varphi)\frac{\partial\varphi}{\partial z}}_{-\frac{\partial\cos(\varphi)}{\partial\varphi}} + \frac{K_1}{J_s}\underbrace{\sin(2\varphi)\frac{\partial\varphi}{\partial z}}_{-\frac{1}{2}\frac{\partial\cos(2\varphi)}{\partial\varphi}} = 0 \quad (3.67)$$

We integrate Eq. (3.67) from $-\infty$ to $z$

$$\int_{-\infty}^{z} \frac{A}{J_s} \frac{\partial}{\partial z} \left[\frac{\partial \varphi}{\partial z}\right]^2 + H_x \frac{\partial \cos(\varphi)}{\partial z} - \frac{K_1}{2J_s} \frac{\partial \cos(2\varphi)}{\partial z} dz = 0 \tag{3.68}$$

and obtain:

$$\frac{A}{J_s}\left(\frac{\partial \varphi(z)}{\partial z}\right)^2 - \underbrace{\frac{A}{J_s}\left(\frac{\partial \varphi(-\infty)}{\partial z}\right)^2}_{0} + H_x \cos(\varphi(z)) - H_x \cos(\varphi(-\infty))$$
$$-\frac{K_1}{2J_s}\cos(2\varphi(z)) + \frac{K_1}{2J_s}\cos(2\varphi(-\infty)) = 0 \tag{3.69}$$

For fields $H_x > \frac{2K_1}{J_s}$ it follows $\varphi(-\infty) = 0$. Hence, we obtain the following differential equation,

$$\frac{A}{J_s}\left(\frac{\partial \varphi}{\partial z}\right)^2 + H_x \cos(\varphi(z)) - H_x - \frac{K_1}{2J_s}\cos(2\varphi(z)) + \frac{K_1}{2J_s} = 0 \tag{3.70}$$

In order to solve Eq. (3.70) we have to take care of the boundary condition according to Eq.(3.63). Due to symmetry we obtained an odd function of the angle $\varphi(z)$ [ Eq. (3.2)}, which is equivalent to

$$m_{x,1} = m_{x,2}$$
$$m_{x,2} = -m_{x,2} \tag{3.71}$$

Hence, we obtain for the general boundary condition at $z = 0$,

$$\begin{pmatrix} \cos\varphi \\ \sin\varphi \\ 0 \end{pmatrix} \times \left[ 2A_1 \begin{pmatrix} \frac{\partial}{\partial z}\cos\varphi \\ \frac{\partial}{\partial z}\sin\varphi \\ 0 \end{pmatrix} - J_{rkky} \begin{pmatrix} \cos\varphi \\ -\sin\varphi \\ 0 \end{pmatrix} \right] = 0 \tag{3.72}$$

which leads to

$$2A_1 \frac{\partial \varphi}{\partial z} = -J_{rkky} \sin(2\varphi) \tag{3.73}$$

We multiply Eq. (3.70) with $A_1$ and evaluate it at the position $z$ = -0, of the interface, where we can insert Eq. (3.73):

$$\underbrace{\left(A\frac{\partial\varphi}{\partial z}\right)^2}_{-\frac{1}{2}J_{rkky}\sin(2\varphi_0)} + AH_xJ_s\cos(\varphi_0) - AH_xJ_s - \frac{AK_1J_s}{2J_s}\cos(2\varphi_0) + \frac{AK_1J_s}{2J_s} = 0 \qquad (3.74)$$

Hence in the limit of $H_x > \frac{2K_1}{J_s}$ we obtain this closed form for the angle $\varphi_0$:

$$\left(\frac{1}{2}J_{rkky}\sin(2\varphi_0)\right)^2 + AH_xJ_s\cos(\varphi_0) - AH_xJ_s - \frac{AK_1}{2}\cos(2\varphi_0) + \frac{AK_1}{2} = 0 \qquad (3.75)$$

Using $H_x = h_x\frac{2K_1}{J_s}$ one obtains

$$\frac{J_{rkky}^2}{4AK_1}\sin^2(2\varphi_0) + 2h_x\cos(\varphi_0) - 2h_x - \frac{1}{2}\cos(2\varphi_0) + \frac{1}{2} = 0 \qquad (3.76)$$

We express for a given angle $\varphi_0$ the corresponding field $h_x$,

$$H_x = \frac{2K_1}{J_s}h_x = \frac{2K_1}{J_s}\frac{1}{4}\frac{\cos(2\varphi_0) - 1 - \overbrace{\frac{J_{rkky}^2}{2AK_1}}^{j_{red}}\sin^2(2\varphi_0)}{\cos(\varphi_0) - 1} \qquad (3.77)$$

$$j_{red} = \frac{J_{rkky}^2}{2AK_1} \qquad (3.78)$$

The angle at the interface $\varphi_0$ as function of the external field $H_x$ for different values of $j_{red}$ that is obtained from the numerical solution of Eq. (3.77) is shown in *Figure 3*.

For sufficiently strong $h_x$ the angle at the interface is exactly $\varphi_0 = 0$. The critical field is reached if this angle starts to deviate from zero, hence if $\varphi_0 > 0$ in Eq. (3.77). Calculating the limit yields for the nucleation field of the domain:

$$H_{x,crit} = \frac{2K_1}{J_s} h_{x,crit} = \frac{2K_1}{J_s} \lim_{\varphi_0 \to 0} \frac{1}{4} \frac{\cos(2\varphi_0) - 1 - \frac{J_{rkky}^2}{2AK_1} \sin^2(2\varphi_0)}{\cos(\varphi_0) - 1} = \frac{2K_1}{J_s}\left(1 + \frac{J_{rkky}^2}{AK_1}\right) \quad (3.79)$$

The field $h_{x,crit}$ can also be interpreted, when the entire structure is perfectly saturated. Hence it is also the saturation field $h_{x,sat} = h_{x,crit}$.

After calculating the angle $\varphi_0$ for a particular field $h_x$, also the entire domain wall profile, $\varphi(z)$ can be obtained. For the sake of an analytical solution not $\varphi(z)$ but $z(\varphi)$ is calculated. From Eq. (3.70) it follows:

$$\frac{\partial z}{\partial \varphi} = \pm \sqrt{\frac{1}{-\frac{J_s H_x}{A}\cos(\varphi(z)) + \frac{J_s H_x}{A} + \frac{K_1}{2A}\cos(2\varphi(z)) - \frac{K_1}{2A}}} \quad (3.80)$$

The indefinite integral of Eq. (3.80) is

$$\tilde{z}(\varphi) = \int \frac{\partial z}{\partial \varphi} d\varphi = \sqrt{\frac{K_1 \cos(\varphi) + K_1 - H_{ext} J_s}{(K_1 - H_{ext} J_s / 2)(H_{ext} J_s - K_1 \cos(\varphi) - K_1)}}$$
$$\times \frac{\sin(\varphi/2)}{|\sin(\varphi/2)|} \sqrt{A} \tanh^{-1}\left(\cos(\varphi/2)\sqrt{\frac{2K_1 - H_{ext} J_s}{K_1 \cos(\varphi) + K_1 - H_{ext} J_s}}\right) \quad (3.81)$$

Hence, the relation between the angle $\varphi$ and the position $z$ is given by:

$$z(\varphi) = \int_{\varphi_0}^{\varphi} \frac{\partial z}{\partial \varphi} d\varphi = \tilde{z}(\varphi) - \tilde{z}(\varphi_0) \quad (3.82)$$

**Appendix C: Equivalence of 1D finite element code and nodal based finite difference code**

In contrast to finite difference methods, where the derivatives are approximated with finite differences finite element methods use basis function to interpolate the unknown function at the domain. The domain is represented by finite elements. The micromagnetic effective field effective field $H_{eff}$ can be calculated from the total energy $E_{tot}$ by,

$$-\int_{\Omega} J_s \mathbf{H}_{eff} \mathbf{v} dV = \delta E_{tot}(\mathbf{m}, \mathbf{v}) \tag{83}$$

, where the functions *v* are test functions that are one on finite element node and zero at all other finite element nodes. Linear or higher order test functions can be used. The right-hand side of Eq. (83) is the Gâteaux derivative of the total energy $E_{tot}(\mathbf{m})$ in the direction $\mathbf{v}$, which can be calculated by finite element packages such as FENICS [16]. The integration over the test functions allows to write Eq. (83) in a discrete form, using the magnetization $m_j$ at nodes points *j*, as

$$\mathbf{A}_{ij} \mathbf{H}_{eff,j} = \mathbf{F}_{ij} \mathbf{m}_j \tag{84}$$

Calculation the effective field according to Eq. (84) would require the solution of a linear system of equations. Hence, mass lumping can be used to approximate the Matrix $A_{ij}$ with a diagonal form $\tilde{\mathbf{A}}_{ij} = a_{ii} \delta_{ij} \approx \mathbf{A}_{ij}$, that can be simply inverted,

$$\mathbf{H}_{eff,j} \approx \tilde{\mathbf{A}}^{-1} \mathbf{F}_{ij} \mathbf{m}_j \tag{85}$$

Eq. (85) shows that the total energy entirely determines the effective field. In the following we will show that the bulk exchange term of Eq. (3.86) for a spacer layer ($\Omega_s$) is equivalent to a surface energy of the form of Eq. (3.88). In Eq. (3.88) we assume the same volume dimensions of the spacer layer and couple the magnetization on one side of the spacer layer and the opposing side according to Eq. (3.88).

The bulk exchange is given by

$$E_{ex}(\mathbf{m}) = \int_{\Omega_s} A_x \left[ \left( \frac{\partial m_x}{\partial x} \right)^2 + \left( \frac{\partial m_y}{\partial x} \right)^2 + \left( \frac{\partial m_z}{\partial x} \right)^2 \right] + A_y \left[ \left( \frac{\partial m_x}{\partial y} \right)^2 + \left( \frac{\partial m_y}{\partial y} \right)^2 + \left( \frac{\partial m_z}{\partial y} \right)^2 \right]$$
$$+ A_z \left[ \left( \frac{\partial m_x}{\partial z} \right)^2 + \left( \frac{\partial m_y}{\partial z} \right)^2 + \left( \frac{\partial m_z}{\partial z} \right)^2 \right] dV \tag{3.86}$$

Without loss of generality we assume that the spacer layer is parallel to the x-y plane. Hence, we set $A_x=0$ $A_y = 0$ and assume no volume nodes in the spacer region. Since we assume linear basis function for the discretion of *m*, the gradient of *m* is constant within each finite element. Hence, we can replace the integral in z direction just by $\Delta z$. From Eq. (3.86) we obtain

$$E_{ex}(\mathbf{m}) = E_{ex}(\mathbf{m}_1, \mathbf{m}_2) = \int \frac{A_z}{\Delta z^2}\left[\left(m_{x,2} - m_{x,1}\right)^2 + \left(m_{y,2} - m_{y,1}\right)^2 + \left(m_{z,2} - m_{z,1}\right)^2\right]\underbrace{\Delta z dA}_{dV} \quad (3.87)$$

, where $\mathbf{m}_1$ is the magnetization as function of space on one side of the spacer and $\mathbf{m}_2$ on the opposing side (Figure 12). We aim to set the exchange parameters $A_x$, $A_y$ and $A_z$ in a way to reach

$$E_{ex}(\mathbf{m}_1, \mathbf{m}_2) = E_{rkky}(\mathbf{m}_1, \mathbf{m}_2) = -\int J_{rkky} \mathbf{m}_1 \mathbf{m}_2 dA \quad (3.88)$$

Considering

$$m_{x,i}^2 + m_{y,i}^2 + m_{z,i}^2 = 1 \quad (3.89)$$

Leads to

$$E_{ex}(\mathbf{m}) = \int \frac{A_z}{\Delta z}[2 - 2\mathbf{m}_1 \mathbf{m}_2] dA \quad (3.90)$$

Rescaling the energy leads to

$$E_{ex}'(\mathbf{m}) = -\int \underbrace{\frac{2A_z}{\Delta z}}_{J_{rkky}} \mathbf{m}_1 \mathbf{m}_2 dA \quad (3.91)$$

Comparing (3.88) and (3.91) leads to

$$J_{rkky} = \frac{2A_z}{\Delta z} \quad (3.92)$$

$$A_z = \frac{J_{rkky} \Delta z}{2} \quad (3.93)$$

In the following we use the bulk exchange of the spacer layer and the derived condition Eq. (3.93) in order to calculate the effective field at the node points of the spacer layer.

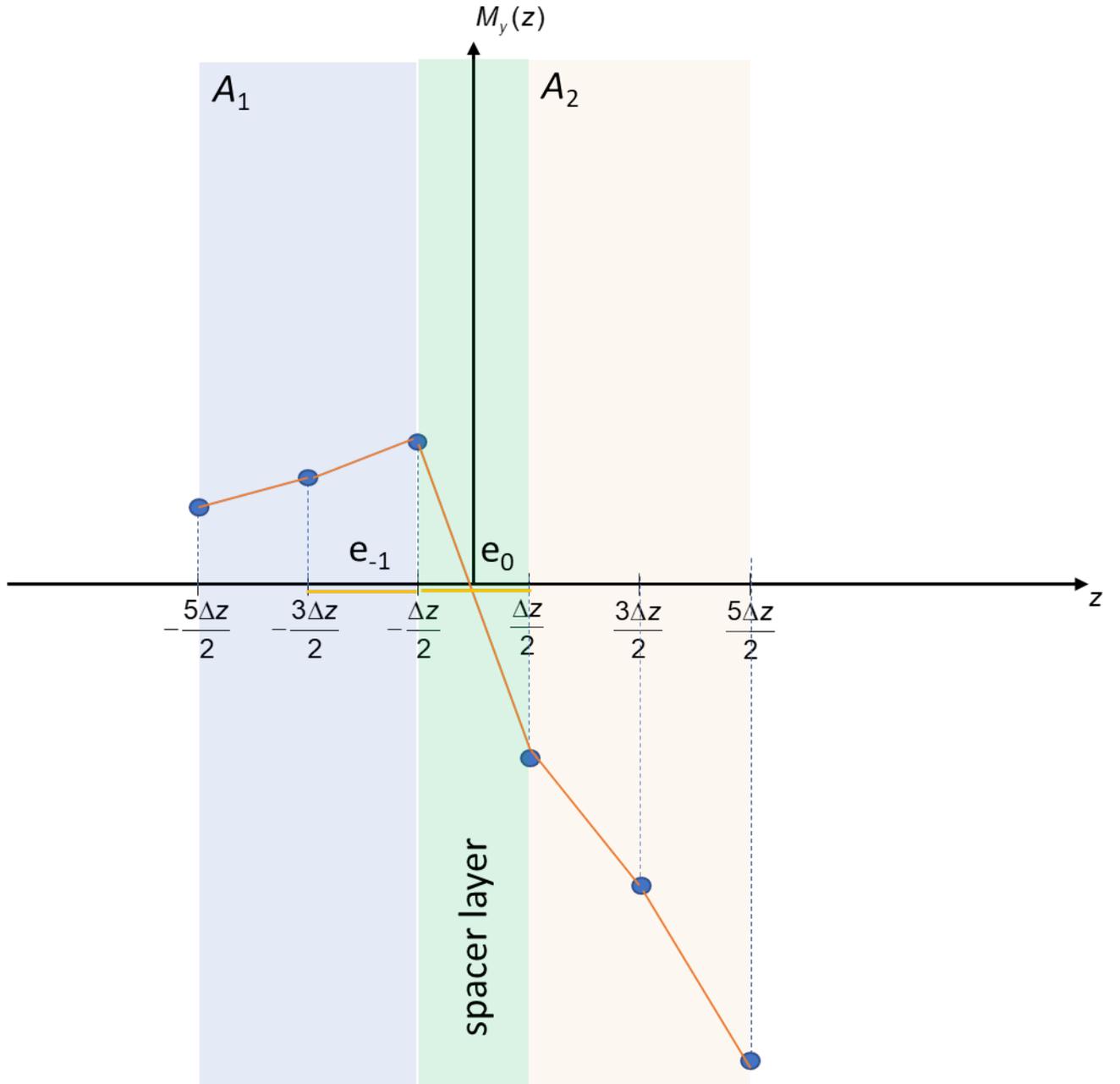

Figure 12: Finite element representation of RKKY coupling

In order to simplify the notation we introduce

$$m_{x,i-1} = m_x(-3\Delta z/2)$$
$$m_{x,i} = m_x(-\Delta z/2)$$
$$m_{x,i+1} = m_x(\Delta z/2)$$

Since linear basis function are used, the gradient within the finite elements is constant. Hence, one can write for the total energy of all terms including the magnetization $m_{x,i}$,

$$E_{ex} = F \int_{e_{-1}} \frac{A_1}{\Delta z^2} \left[ (m_{x,i} - m_{x,i-1})^2 + (m_{y,i} - m_{y,i-1})^2 + (m_{z,i} - m_{z,i-1})^2 \right] dz +$$
$$F \int_{e_0} \frac{A_{rkky}}{\Delta z^2} \left[ (m_{x,i+1} - m_{x,i})^2 + (m_{y,i+1} - m_{y,i})^2 + (m_{z,i+1} - m_{z,i})^2 \right] dz + \ldots \quad (3.94)$$

, where $F$ is the area in the x and y direction. The mass lumped equation for the effective field according to Eq. (85) can be written in the form of the Box scheme [17] with the corresponding volume $V_i = F\Delta z$ for each nodal point $i$ as,

$$H_{ex,x}(-\Delta z/2) = -\frac{1}{V_i J_{s,average}} \frac{\partial E_{ex}}{\partial m_{x,i}} = -\frac{1}{F J_s \Delta z/2} \frac{\partial E_{ex}}{\partial m_{x,i}} =$$
$$-\frac{1}{J_s \Delta z/2} \frac{\partial}{\partial m_{x,i}} \left( \frac{A_{rkky}}{\Delta z} \left[ (m_{x,i+1} - m_{x,i})^2 + (m_{y,i+1} - m_{y,i})^2 + (m_{z,i+1} - m_{z,i})^2 \right] \right) + \quad (3.95)$$
$$-\frac{1}{J_s \Delta z/2} \frac{\partial}{\partial m_{x,i}} \left( \frac{A_1}{\Delta z} \left[ (m_{x,i} - m_{x,i-1})^2 + (m_{y,i} - m_{y,i-1})^2 + (m_{z,i} - m_{z,i-1})^2 \right] \right)$$

Since, the saturation magnetization in the spacer layer is zero $J_{s,average} = J_s/2$. For the exchange field one gets:

$$H_{ex,x}(-\Delta z/2) = -\frac{1}{F J_s \Delta z/2} \frac{\partial E_{ex}}{\partial m_{x,i}} = \frac{2}{J_s \Delta z^2} \left( 2 A_{rkky} m_{x,i+1} + 2 A_1 m_{x,i-1} \right) \quad (3.96)$$

Using the proper value of $A_{rkky}$ to represent $J_{rkky}$ according to Eq. (3.93) one gets.

$$\mathbf{H}_{ex}(-\Delta z/2) = \frac{2}{J_s \Delta z^2} \left( J_{rkky} \Delta z \mathbf{m}_{i+1} + 2 A_1 \mathbf{m}_{i-1} \right) \quad (3.97)$$

which is equivalent to the derived equation for the nodal finite difference formula according to Eq. .